# Electric Vehicle Battery Remaining Charging Time Estimation Considering Charging Accuracy and Charging Profile Prediction

Junzhe Shi, Min Tian*, Sangwoo Han, Tung-Yan Wu, Yifan Tang

SERES, Santa Clara, CA, 95054, USA

* Corresponding author: SERES EV, Santa Clara, CA, 95054, USA. E-mail address: mintian89@gmail.com (Min Tian)

**ABSTACT**

Electric vehicles (EVs) have been growing rapidly in popularity in recent years and have become a future trend. It is an important aspect of user experience to know the Remaining Charging Time (RCT) of an EV with confidence. However, it is difficult to find an algorithm that accurately estimates the RCT for vehicles in the current EV market. The maximum RCT estimation error of the Tesla Model X can be as high as 60 minutes from a 10 % to 99 % state-of-charge (SOC) while charging at direct current (DC). A highly accurate RCT estimation algorithm for electric vehicles is in high demand and will continue to be as EVs become more popular. There are currently two challenges to arriving at an accurate RCT estimate. First, most commercial chargers cannot provide requested charging currents during a constant current (CC) stage. Second, it is hard to predict the charging current profile in a constant voltage (CV) stage. To address the first issue, this study proposes an RCT algorithm that updates the charging accuracy online in the CC stage by considering the confidence interval between the historical charging accuracy and real-time charging accuracy data. To solve the second issue, this study proposes a battery resistance prediction model to predict charging current profiles in the CV stage, using a Radial Basis Function (RBF) neural network (NN). The test results demonstrate that the RCT algorithm proposed in this study achieves an error rate improvement of 73.6 % and 84.4 % over the traditional method in the CC and CV stages, respectively.

## 1. INTRODUCTION

The number of EVs on the market continues to rise as they play a key role in achieving the world's efforts to reduce the impacts of climate change. Because EVs use off-board electricity generated by renewable energy sources, replacing an internal combustion engine (ICE) vehicle [1] with an EV could greatly reduce carbon dioxide emissions that contribute to climate change and smog [2]. The US government plans to limit the carbon dioxide emissions of passenger cars to 88g/km by 2025 [3]. There are many additional benefits of EVs, including high energy efficiency and instant torque supply availability [4]. With the rising demand for EVs, many studies have focused on EV operational optimization, such as energy management optimization [5][6][7] and battery life optimization [8][9][10]. However, achieving accurate RCT estimates is a prerequisite for optimal operations [11][12]. To fully charge an EV may take several hours. Thus, an accurate RCT estimation algorithm for EVs is crucial in the following three aspects:
1) Personal trip planning [13]: helping people who drive EVs arrange their trip schedule,
2) Public EV charging station planning [14]: allowing each EV to wait in an optimally-selected queue to reduce the total waiting times of all the waiting EVs that may be backed up due to the lack of charging infrastructure,
3) Fleet management for EV [15]: optimizing the scheduling of real-time EV fleet operation, including relocation and charging. For example, an EV fleet can gain optimal scheduling and charging costs considering electricity price variations.

Although an accurate RCT estimation algorithm is gaining in importance, there seems to be very limited information in the literature regarding studies to develop an accurate RCT estimate algorithm. Within the industry, finding information on the issue of battery RCT estimation is also a challenge. A 2016 Tesla Model X 90D (90kwh battery pack) with software version 2020.40.9.2 (latest at Dec 1, 2020) was used for supercharging with a 150-kW maximum charger power. As shown in Fig. 1, while charging from 10 % to 99 % SOC, the RCT estimation error of the Tesla Model X system is extremely high. The RMSE of the RCT estimate is 42.1565 mins, and the highest estimate is 61 minutes.

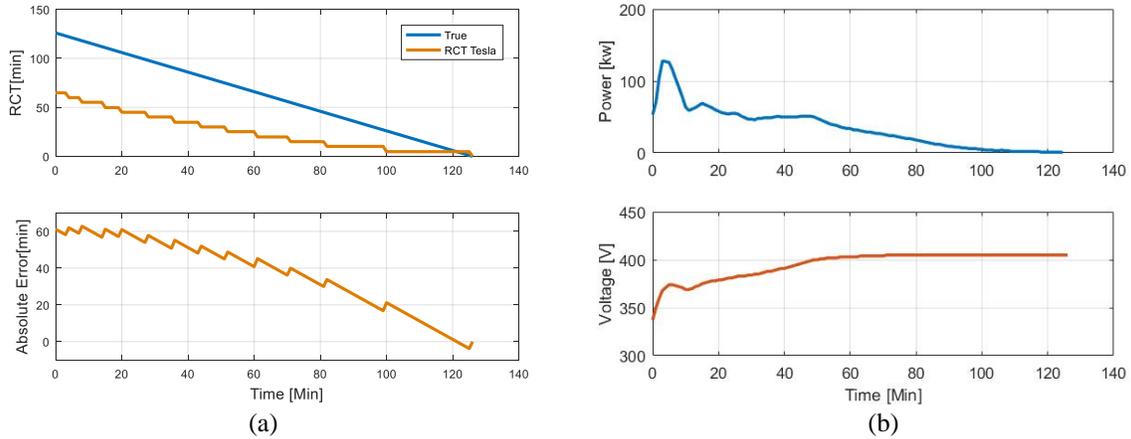
Fig. 1. Charging a Tesla Model X from 10 % to 99 % SOC: (a) the Tesla RCT estimation, and (b) the power and voltage profiles of the charging process.

As shown in Fig. 2, the charging processes of a Li-ion battery can be commonly represented as CC and CV stages separately. In the CC stage, the charging current follows the designed current, with initial charging of a relatively higher current and a finishing rate of low current to avoid excessive gassing, overheating, and battery degradation [16][17]. The CC to CV transition happens when the terminal voltage of the battery first reached the cut-off voltage. In the CV stage, the cell was charged to the cut-off voltage, followed by a CV charging until the current decreases to the cut-off current [18].

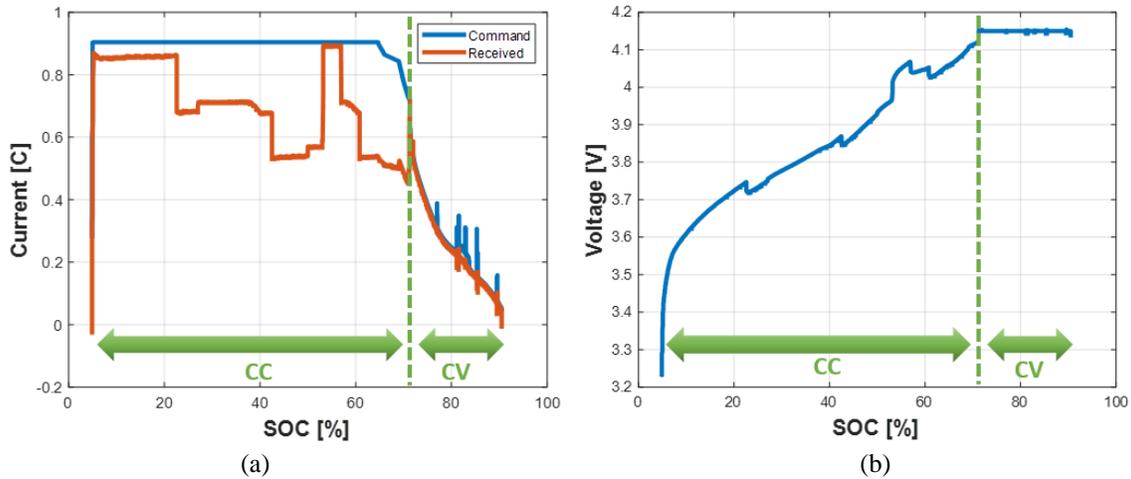
Fig. 2. Voltage profiles in constant current and constant voltage charging processes.

There are two challenges involved with RCT estimation. First, oftentimes and especially during fast charge, commercial chargers cannot provide requested charging currents during the CC stage. Second, it is hard to predict the current profile in the CV stage. The charging accuracy is usually considered as a constant. However, this assumption is not consistent with the real charging processes, in which the efficiency depends on the multiple impact factors. In [19][20][21], the impacts of charger type, charging current, and temperature on the charging efficiency are discussed. However, all of these do not address a method to describe the actual real-time charging accuracy of EVs. In the past five years, battery state estimation using machine learning has gained more researchers' attentions. In [22], a Radial Basis Function (RBF) neural network (NN) is employed for Li-ion battery SOC estimation in electric vehicles. In [23], a probabilistic neural network (PNN) is used to estimate the state of health (SOH) of Li-ion batteries. These data-driven approaches greatly enhance the state of predicting accuracy of complex, nonlinear systems such as Li-ion batteries [24]. In [25], an EV charging profile prediction method is proposed, using an artificial neural network (ANN) that considers the previous charging profiles, initial SOC, and final SOC. However, they do not sufficiently consider the impact factors for the current battery charging predictions, such as battery resistance variations due to the aging effect. The current profile during the CV charge is changing because battery resistance increases as a battery ages [26].

In the study described in this paper, a novel battery RCT estimation method is proposed, by considering charging accuracy in the CC stage and the charging profile in the CV stage. As shown in Fig. 3, the proposed method takes into consideration the inputs of SOC, starting SOC, target SOC, battery temperature, command and received current, and charger type. For the CC stage, by analyzing the charging process using real data from charging stations and EVs, the charging accuracy prediction algorithm is proposed by considering the confidence interval between the historical charging accuracy and real-time charging accuracy data. The impacted factors (SOC, starting SOC, and internal temperature) of the Li-ion battery resistance in the CV stage are tested and analyzed. An RBF NN model is developed to predict the resistance of a Li-ion battery for current profile prediction with given OCV and cut-off voltage in the CV stage while considering the battery degradation effect. The primary contributions of the study in this paper are:
1) A novel battery RCT estimation algorithm developed for EVs,
2) A method for charging accuracy updating,
3) A model of current profiles and battery resistance prediction using RBF NN.

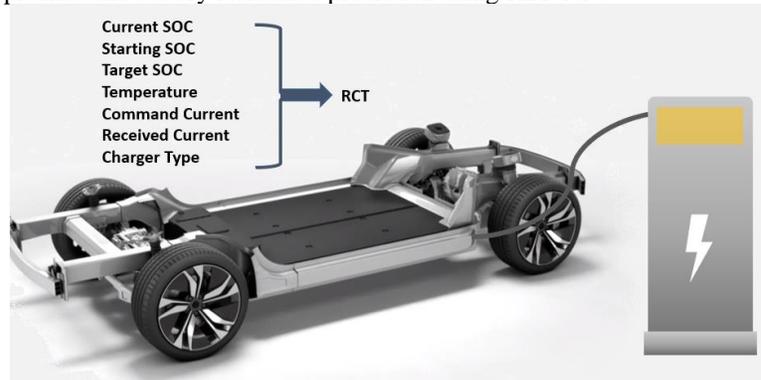

Fig. 3. The schematic diagram of the Remaining Charging Time estimation with the proposed method of an electric vehicle.

The paper is organized as follows. In section 2, the proposed method of RCT estimation is formulated and presented. The sensitivity study of battery resistance in the CV stage is discussed in section 3. In Section 4, the comparison study is performed to evaluate the performance of the proposed method. Section 5 provides a summary and suggests future work.

## 2. ALGORITHM FOR BATTERY REMAINING CHARGING TIME

This section introduces and discusses the algorithms proposed in this study for the battery RCT estimation for CC and CV charging processes, respectively.

The current SOC, starting SOC, and target SOC are defined as sequentially the current real-time estimated SOC, the starting SOC at the beginning of the charging, and the target SOC which is targeted to get at the end of the charging. Besides, the term CV turning SOC represents the SOC value when the CC stage transitions to the CV stage. The starting SOC and the target SOC can reside in either the CC range or CV range. Therefore, three different RCT estimation scenarios need to be discussed accordingly, and the key point is to determine the starting and ending SOC of CC and CV stages separately. As shown in Fig. 4, for the first scenario (Scenario 1), the current and target SOC are all in the CC range, in which only the calculation of the CC part is needed. The starting and ending points for the CC charging process are the current SOC and target SOC. For the second scenario (Scenario 2), the current SOC is in the CV range, which means only the CV charging process needs to be incorporated into the calculation. The starting and ending points for the CV charging process are the current SOC and target SOC. For the third scenario (Scenario 3), the current SOC is in the CC range, and the target SOC is in the CV range, which consists of both CC and CV charging processes. The starting and ending points for the CC charging process are the current SOC and CV turning SOC, while the starting and ending points for the CV charging process are the CV turning SOC and target SOC.

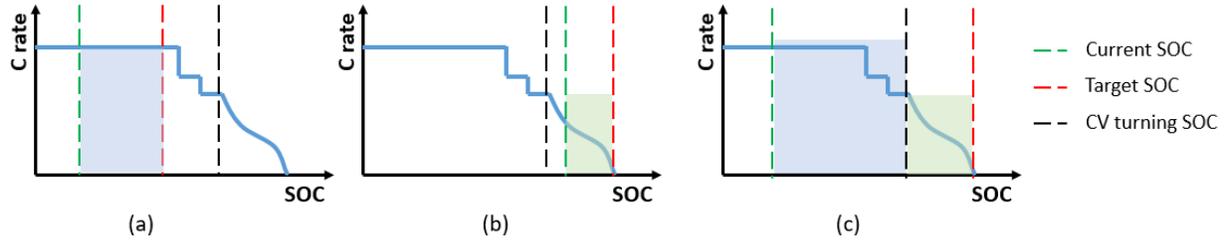

Fig. 4. Three charging scenarios of RCT estimations: (a) Scenario 1, current SOC and target SOC are all in the CC range, (b) Scenario 2, current SOC is in the CV range, and (c) Scenario 3, current SOC is in the CC range, and target SOC is in the CV range.

## 2.1 REMAINING CHARGING TIME ESTIMATION IN THE CC STAGE

In the CC stage, the designed current profile is known. The problem with the RCT estimation is that the received current from the charger does not always match the command current. Thus, a charging accuracy in the CC range, $\eta_{cc}$, is introduced to represent the difference between the received and command current.

In the CC stage, the current in the designed current profile is not always a constant. To reduce battery degradation and ensure the safe operation, the command current decreases as the SOC increases. Thus, the SOC range of the CC charging process is partitioned, based on the designed current profile, to ensure that the command current is a constant value in each partition. After partitioning the SOC range of the CC charging process in N segments, so based on the designed current profiles. As shown in Fig. 5, the RCT in the CC stage can be calculated by,

$$RCT_{CC} = \sum_{i=0}^{N} \frac{\Delta SOC(i)}{C_{rate}(i)} \eta_{cc} \qquad (1)$$

where N is numbers of partitions, the $\Delta SOC$ is the delta SOC of each partitioned range, and $C_{rate}$ is the C-rate of the designed current in each partition. N is determined by the designed current charging profile and ending SOC, which allows each partition to have a constant commanded current.

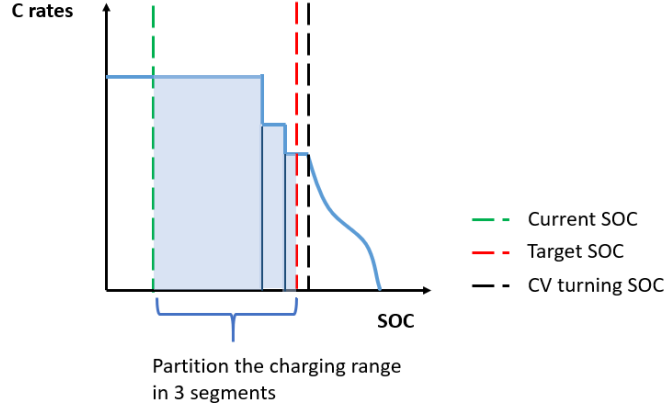

Fig. 5. The SOC partitions of the constant current charging process for Scenario 1.

The charging accuracy in the CC stage, $\eta_{cc}$, which is initialized by the pre-study knowledge about the charger type and its correlated historical charging accuracy, is updated real-time by the exponential moving average method [27] with the instantaneous charging accuracy, $\eta_{cc\,instant}$, and previous charging accuracy,

$$\eta_{cc\,instant}(k) = \frac{I_{received}(k)}{I_{commanded}(k)} \qquad (2)$$

$$\eta_{cc}(k+1) = \left(1 - \alpha_{update}(k)\right) * \eta_{cc}(k) + \alpha_{update}(k) * \eta_{cc\,instant}(k) \qquad (3)$$

where $I_{received}$ and $I_{commanded}$ are the received and command current, $\eta_{cc\,instant}$ is the instantaneous charging accuracy, $\eta_{cc}$ is the updated charging accuracy, and $\alpha_{update}$ is a time-variant updating rate. The updating rate, $\alpha_{update}$, dynamically varies by considering the confidence coefficient between the historical and process efficiencies,

$$\alpha_{update} = \alpha_{slow} + (\alpha_{fast} - \alpha_{slow})e^{\frac{SOC_{current}-SOC_{target}}{SOC_{current}-SOC_{start}}} \tag{4}$$

where $SOC_{start}$, $SOC_{current}$, and $SOC_{target}$ are the starting, current, and target SOC values of a charging process, and $\alpha_{fast}$ and $\alpha_{slow}$ are two designed updating rates. The value of $\alpha_{fast}$ is higher than the value of $\alpha_{slow}$. A higher value of update rate means a faster updating process that puts more weight on the instantaneous charging accuracy than the previous charging accuracy. At the beginning of the charging, the historic charging accuracy is more accurate in representing the charging accuracy of the whole charging process than the instantaneous charging accuracy. Thus, the value of the $\alpha_{update}$ should be small. Namely, when $SOC_{current}$ is close to $SOC_{start}$, $\alpha_{update}$ should be closer to $\alpha_{slow}$. While the current SOC is closer to the target SOC, instantaneous charging accuracy better represents the charging accuracy of the rest of the charging process than before. Thus, a faster updating rate should be taken into the calculation in this condition. Namely, when $SOC_{current}$ is closer to $SOC_{target}$, $\alpha_{update}$ should be closer to $\alpha_{fast}$.

## 2.2 REMAINING CHARGING TIME ESTIMATION IN THE CV STAGE

During the charging process in the CV stage, the command current needed to sustain the desired constant voltage is relatively small. Therefore, the received current can follow the command current very closely, even the charger in a derating mode, which frequently happens in the CC stage. However, the RCT estimation challenge in the CV stage is that there is not a previously known current profile. Thus, to estimate the RCT in the CV stage, there is a need to predict current profile accurately. In the CV charging process, the terminal voltage of the battery is fixed at the cut-off voltage, and the charging current decreases slowly, which leads to slow dynamic changes of the battery. Thus, as shown in Fig. 6 (a), a Rint model of a battery is used in this study to represent the electrical behaviors of a battery in the CV charging process. The battery model comprises an open-circuit voltage (OCV) and an ohmic resistor.

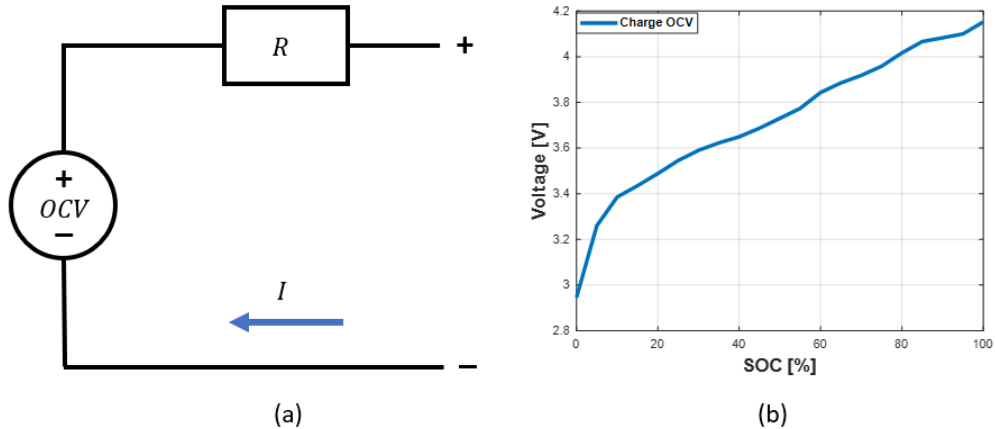

(a) (b)

Fig. 6. Equivalent circle model of a battery: (a) the structure of the Rint model, and (b) the curve of SOC vs. OCV.

According to the Rint model of a battery, the CV charging C-rate can be calculated by,

$$C_{rate} = \frac{V_T - OCV}{R * C_p} \tag{5}$$

where $I$ is the charging current, $V_T$ is the terminal voltage which is equal to the cut-off voltage of the battery in the CV charging process, OCV is the open-circuit voltage, $R$ is the ohmic resistance, and $C_p$ is the capacity of the battery. As shown in Fig. 6 (b), the value of the OCV is a known function of the SOC. However, the resistance is an unknown value that depends on different impact factors. Basing on the sensitivity study in Section 3, the current SOC, starting

SOC of the charging process, and the internal temperature of a Li-ion battery are selected as the dominant impact factors used to predict the battery resistance in the CV stage.

Due to the high nonlinearity of the resistance, an RBF NN [28] is selected in this study to predict the resistance of a battery under given conditions. RBF NN is a three-layer feedforward neural network using radial basis functions as activation functions. Fig. 7 illustrates the structure of the RBF NN model used in this study, which contains three layers. The input layer corresponds to the inputs of the network. The hidden layer consisting of a number of RBF non-linear activation units. The output layer corresponds to the final output of the network.

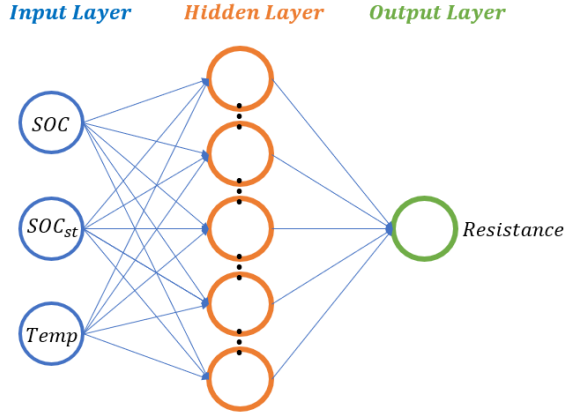

Fig. 7. RBF-NN model for forecasting the resistance of a li-ion battery during the constant voltage charging process.

To perform nonlinear transforms, the gaussian function is used as the radial basis function in the hidden layer,

$$H_i = e^{-\frac{1}{2\sigma_i^2}\|X-C_i\|^2} \quad (6)$$

where $H_i$ is a radial basis function in the hidden layer, the $C_i$ is the center of radial basis function, $\sigma_i^2$ is the spread width, and $X$ is the input vector. During the training, the center vector $C_i$, constituted by SOC, starting SOC, and temperature, of the RBF NN in the hidden layer, is selected using k-means clustering [29] with a large amount of test data.

In the calculation of neurons in the output layer, a linear activation function is adopted to predict the resistance in the given condition,

$$R = \sum_{i=1}^{N} W_i H_i \quad (7)$$

where $R$ is the predicted resistance, $N$ is the numbers of nodes in the hidden layer, $H_i$ is a radial basis function in the hidden layer, and $W$ is the optimal weight vector. In the training processes, the values of the weight vector $W$ are determined by fitting the linear model [30] concerning outputs of the hidden layer and the mean squared error (MSE) between predicted voltage and actual measured voltage. The MSE is calculated by,

$$MSE = \frac{1}{N}\sum_{i=1}^{N}\left(V_{predicted_i} - V_{measured_i}\right)^2 \quad (8)$$

$$V_{predicted_i} = OCV_i + R_i I_i \quad (9)$$

where $N$ is the numbers of the training data, OCV is the real-time estimated open-circuit voltage, $R$ is the real-time predicted resistance, $I$ is the real-time measured charging current, $V_{predicted_i}$ and $V_{measured_i}$ are the predicted and measured terminal voltage of the battery.

Since the statistical distributions of the input data do not change with time, the centers of RBF units are fixed once they have been trained offline. However, as a battery ages, its resistance will rise. Different operating conditions of a battery, such as different driving behaviors of drivers, lead to different aging results of batteries. Thus, the resistances of a battery at the CV stage should be calculated and stored for updating the output weights of the RBF NN and adopting the aging of batteries. During the online training processes, the output weight vector, $W$, is updated with the gradient descent algorithm [31],

$$W := W - \alpha \frac{1}{m} \sum_{i=1}^{m} \frac{\partial\left(\left(V_{predicted_i} - V_{measured_i}\right)^2\right)}{\partial W} e^{-t_i} \tag{10}$$

Where $\alpha$ is the learning rate, $m$ is the numbers of stored data, $V_{predicted_i}$ and $V_{measured}$ are the predicted and measured terminal voltage of the battery, and $t$ represents the serial number of stored data during DC fast charging. The serial number of the stored data is used to indicate if the data is new or old. Because the resistance of a battery rises while a battery ages, the new data is more representative of the current characteristics of the battery. Thus, new data always has a smaller serial number than old data, which helps to assign higher weights on the new data than the old data during the training. When data is too old, with a serial number higher than a designed threshold, the data will be discarded to save storage space and reduce computation cost.

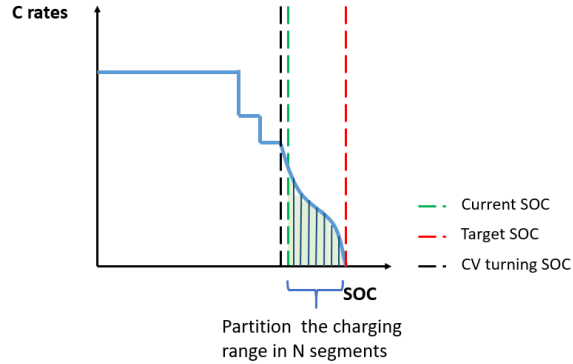

Fig. 8. The SOC partitions of the constant current charging process for Scenario 2

After obtaining the resistances of battery in the CV state by applying RBF NN model, the charging current can be determined by Eq. (11). As shown in Fig. 8, the charging stage is partitioned in N segments, each with an interval of 1% SOC. The RCT in the CV stage can be calculated by,

$$RCT_{CV} = \sum_{i=0}^{N} \frac{\Delta SOC(i)}{C_{rate}(i)} \eta_{cv} \tag{11}$$

Because the received current follows the commanded current very well in CV stage, the charging accuracy factor, $\eta_{cv}$, is assumed as 100%.

## 3. SENSITIVITY STUDY
In this section, a sensitivity study of Li-ion battery resistance in the CV stage is discussed. The Li-ion cells used in the experiments are commercial 21700 cylindrical cells (NCA/Gr-Si blend) with a nominal capacity of 4.8 Ah under 1 C discharge at 25 °C. All Li-ion cells were tested with Arbin Instruments LBT21084 battery cycler and inside the TestEquity 123C chamber for temperature control. The constant current constant voltage (CCCV) charging method was applied to Li-ion cells. During the charging period, the cell internal resistance was computed based on Rint model,

$$R = \frac{V_T - OCV}{I} \tag{12}$$

where $R$ is the resistance, $V_T$ is the terminal voltage, OCV is the open-circuit voltage, and $I$ is the charging current. In the tests, both 0.3 C and 0.7 C C-rate were applied for the constant current charging mode. Then the CC mode was switched to the CV mode once cell voltage reached the cut-off voltage and held at that voltage until the current fell below C/20. The CCCV charging started from various SOC under 25, 35, and 50 °C ambient temperatures. The initial SOC was set based on Coulomb counting under 25 °C ambient temperature.

The cell internal resistances computed during the CCCV charging period are plotted in Fig. 9 (a). As observed in Fig. 9 (a), the cell resistance rises rapidly at the beginning of the CC charging process. This rapid rise of resistance in the beginning is primarily due to concentration polarization build-up in the electrolyte phase [32]. Once the Li$^+$ concentration gradient stabilizes, the rate of concentration polarization increase slows. In addition, Fig. 9 (a) shows that the cell resistance is lower under higher C-rate during the CC charging process. The lower cell resistance under higher C-rate is likely due to lower kinetic resistance or nonlinear relationship between the concentration polarization and the applied current [33]. For visual clarity, only the cell resistances during CV charging are plotted separately in Fig. 9 (b). As shown in Fig. 9 (b), the SOC, where the CC to CV transition takes place, is dependent on the initial SOC, C-rate, and cell temperature. Also observed in Fig. 9 (b) is that towards the end of the CV charging phase, the cell resistance increases exponentially. During the CV charging phase, the cell voltage is held constant, and the cell open circuit potential (OCP) converges to the fixed cell voltage. According to the Butler-Volmer kinetics, as the surface overpotential approaches zero, the resulting reaction current exponentially decreases. That is, as the surface overpotential approaches zero during the constant voltage mode, the resulting reaction current decreases at a much faster rate. This causes the computed cell resistance to rise during the CV mode. Cell temperature is another significant factor that affects the cell internal resistance. The calculated battery resistance during the CV mode under different ambient temperature is plotted in Fig. 10 to illustrate the effect. In Fig. 10, the initial SOC and the applied C-rate during the CC mode are kept the same. As the cell internal resistance generally decreases with increasing temperature [34], the SOC at which the CC mode transitions to the CV mode is higher with increasing temperature. Moreover, the overall battery resistance during the CV mode is lower with increasing temperature. In summary, the above demonstrates that the SOC, initial SOC, and the battery temperature have an impact on battery resistance during the CV mode.

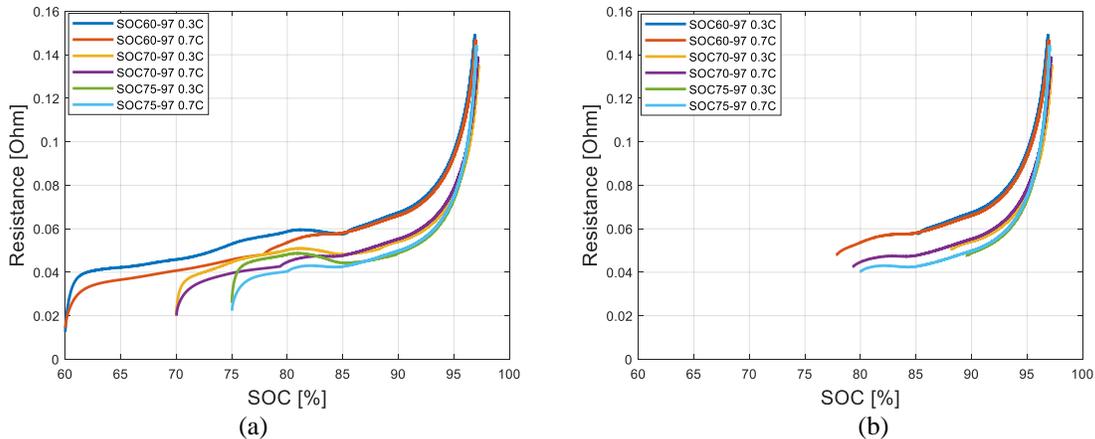

Fig. 9. The effects of starting SOC and C-rate in the constant current stage on battery resistances at 25 degrees centigrade: (a) resistances in the constant current to constant voltage stage, and (b) resistances in the constant voltage stage only.

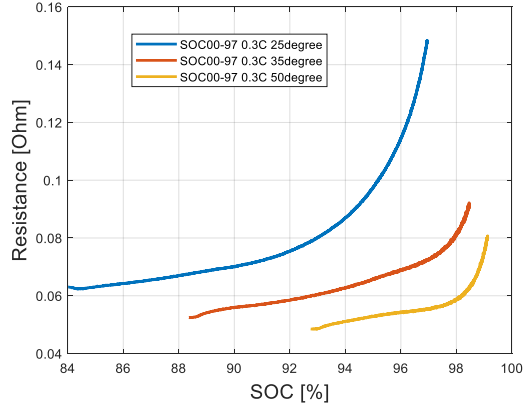

Fig. 10. Battery resistances in the constant voltage stage under 25 °C, 35 °C, and 50 °C ambient temperatures.

## 4. PERFORMANCE DISCUSSION

As discussed in the above sections, there are two charging processes, CC and CV. In this section, the proposed method is verified and discussed by testing the CC charging case and the CV charging case. Meanwhile, a traditional method commonly used in the electric automobile industry is introduced as a benchmark for the proposed method.

The traditional RCT estimation method uses charging current, battery capacity, and delta charging SOC as the inputs. The equation of the RCT calculation used in the conventional RCT algorithm is,

$$RCT = \frac{\Delta SOC * Cp}{Min(I_{charger}, I_{command})}\eta \qquad (13)$$

where $\Delta SOC$ is the delta charging SOC which is equal to the difference between the target SOC and starting SOC, $Cp$ is the battery capacity, $I_{charger}$ is the maximum supply current of the charger, $I_{command}$ is the command charging current, and $\eta$ is the constant pre-set charging accuracy, which is set to be 90% and 100% in the CC and CV scenarios, respectively.

### 4.1 SYSTEM CONFIGURATIONS

As shown in Fig. 11, a SERES SF5 electric vehicle is used for the experimental tests. The test vehicle is equipped with a 90-kWh Li-ion battery pack and with a maximum powertrain output power of 510 kW. The RCT estimation is also affected by the type and charging power of electric vehicle supply equipment (EVSE). The charging type and level are based on the power distribution standards and maximum power. In this study, a DC fast charging supplier refer to as GB/T 20234 Standards which employed in China is used.

In Fig. 11, the test vehicle conducts charging by connecting to a DC fast charging station in Chongqing, China. It is a level 2 DC EVES with a maximum charging power of 300 kW and maximum current of 250 Amps.

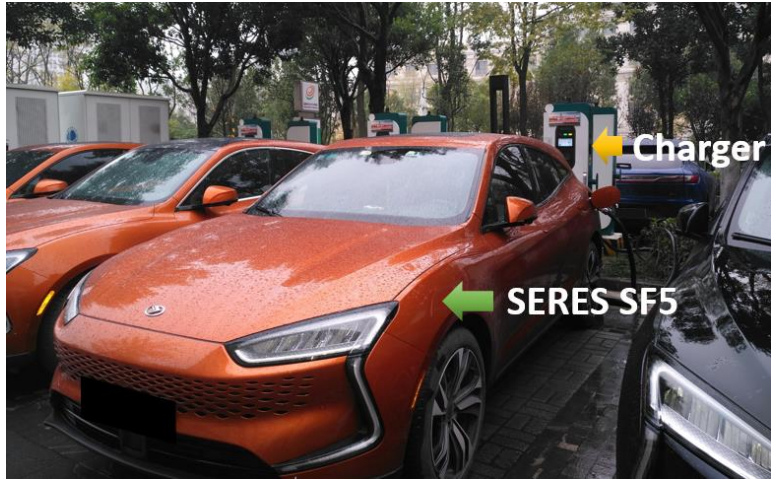

Fig. 11. Test the performance of the Remaining Charging Time estimation algorithm with a DC fast charger and SERES SF5 electric vehicle.

## 4.2 PERFORMANCE IN THE CC STAGE

This section will present the test result of the CC charging process, in which the batteries were charged from 5% to 70% SOC. As shown in Fig. 12, the real charging current varies regardless of a designed smooth charging profile, which is due to the unstable operation of the electricity grid, charging station, and various operational control and management within the battery system. The unpredictable charging current leads to difficulties in the RCT estimation.

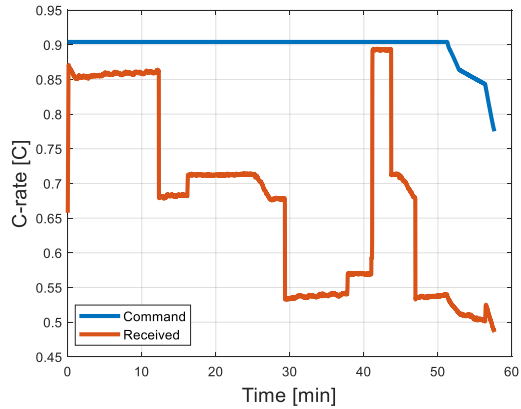

Fig. 12. Command and received C-rates in the constant current charging scenario.

The proposed method is compared with the traditional RCT estimation method to evaluate its performance. As shown in Fig. 13, the estimation results of proposed and traditional methods are compared with the true RCT. The root mean squared error (RMSE) of each method is gained by using,

$$RMSE = \sqrt{\frac{1}{m}\sum_{i=1}^{m}\left(RCT_i - \widehat{RCT_i}\right)^2} \qquad (14)$$

Where $\widehat{RCT}$ and $RCT$ are the estimated and true values of the remaining charging time, and $m$ represents the total number of the estimations in the whole charging process. The RMSEs of the traditional and proposed methods are 7.6288 and 2.0165 minutes, respectively. In the test, the overall charging accuracy of the charger is 0.748. With the information of the charger type and its correlated historical charging accuracy, the charging accuracy of the proposed method is initialized as 0.7, which is 6.4 % different from the actual overall charging accuracy and leads to a small estimated error at the beginning. On the other side, the traditional method does not have any pre-knowledge about the charging accuracy of the charger and always uses a constant charging accuracy, 0.9, which causes a significant estimation error. As a result, the proposed method introduces a 73.57% improvement over the traditional method.

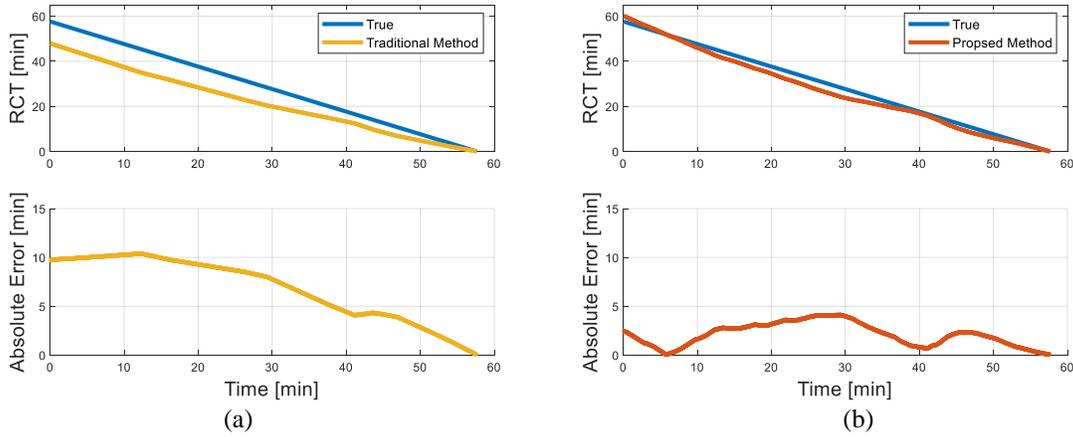

Fig. 13. The remaining charging time estimation results in the CC charging stage: (a) traditional method, and (b) the proposed method with a relatively accurate initial charging accuracy.

Having good pre-knowledge about the charger plays an important role in an RCT estimation. However, in some extreme cases, the pre-knowledge of a charger may be unavailable or even incorrect. To further validate the robustness of the proposed method, the charging accuracy of the proposed method is initialized by an incorrect value, 0.5, which leads to a 33.15 % error than the actual whole charging accuracy, 0.748. As shown in Fig. 14 (a), the confidence coefficient of instantaneous charging accuracy is increasing, while the current SOC is closer to the target SOC. A higher confidence coefficient leads to a higher updating rate, which puts more weight on the instantaneous charging accuracy than the previous charging accuracy. Thus, the charging accuracy applied in the proposed algorithm gradually eliminates the harmful effect of the wrong charging accuracy initialization. The estimation error is decreasing while SOC is approaching the target SOC and is converging to zero at the end of the charging. As shown in Fig. 14 (b), after 12 minutes of charging, the estimation error drops to 10 minutes. By leveraging the exponential moving average method, the estimation time is always smooth. Even the received C-rate current jumps from 0.6 C to 0.9 C at 41 minutes.

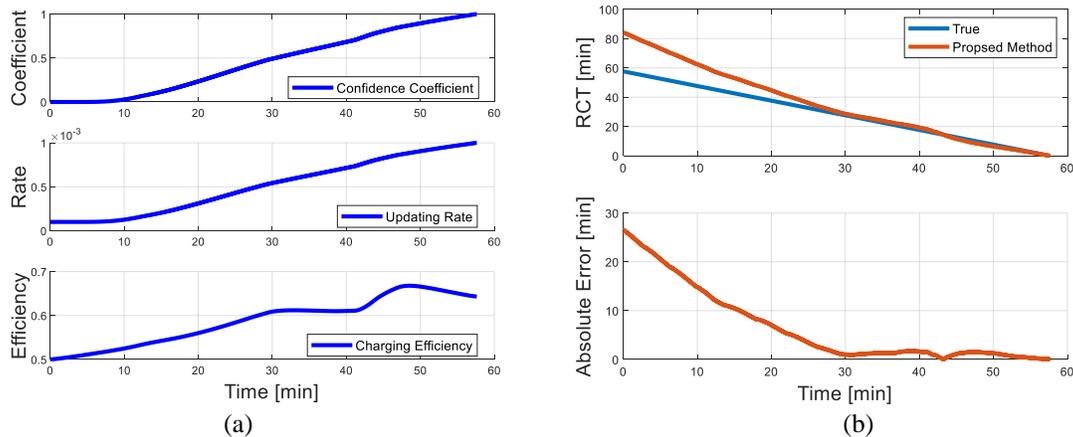

Fig. 14. The remaining charging time estimation results of the proposed method in the CC charging stage with a wrong initial charging accuracy

### 4.3 PERFORMANCE IN THE CV STAGE

The proposed and traditional methods were tested in the CV stage, in which the batteries were charged from 71% to 90% SOC. As shown in Fig. 15, the received current can almost follow the command current. However, the command current is not always smooth, due to the additional energy consumptions of the vehicle besides charging the batteries, such as the air conditioning compressor consumption. When SOC is higher than 83%, the resistance of a battery increases exponentially as the value of SOC is closer to 100%. In addition, the OCV of a battery also increases with the SOC of the battery. Thus, based on equation (5), in the CV stage, the charging current becomes smaller while the current SOC is closer to the target SOC.

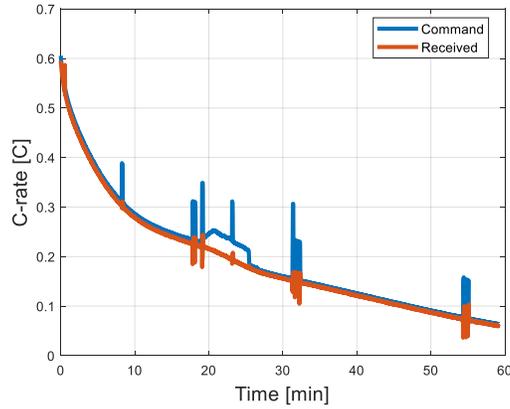
Fig. 15. Command and received C-rates in the constant voltage scenario.

As shown in Fig. 16, the estimation results of the traditional and proposed method are compared with the true RCT. Due to the lack of information on the future charging current, the traditional method always leads to an underestimation of the RCT. Besides, a fluctuating command current also results in an undesired fluctuation of the estimated RCT of the traditional method, which causes an unfavorable user experience. For the proposed method, with the assistance of the resistance prediction, the future current profile can be obtained. As shown in Fig. 17, with a well-trained RBF NN resistant prediction model, the predicted resistances are close to the true resistances of the tested battery pack, which provides an accurate and smooth predicted current profile. The underestimation phenomenon is shown in the RCT estimation of the proposed method. While a battery ages, its resistance increases as well. With the same operating condition, the resistances in the earlier stored data are lower than the resistances in the new data. Thus, during the online training of the resistance prediction model, the earlier stored resistance data leads to an underestimation of the resistance. The underestimated resistances cause an overestimated current profile, which results in an underestimated RCT. However, by taking the new data and assigning smaller weights to the old data, the slight underestimation of the RCT of the proposed method is acceptable. As a result, the RMSE of the traditional and proposed method are 17.534 and 2.735 minutes, respectively. The proposed method produces an 84.4% improvement over the traditional method.

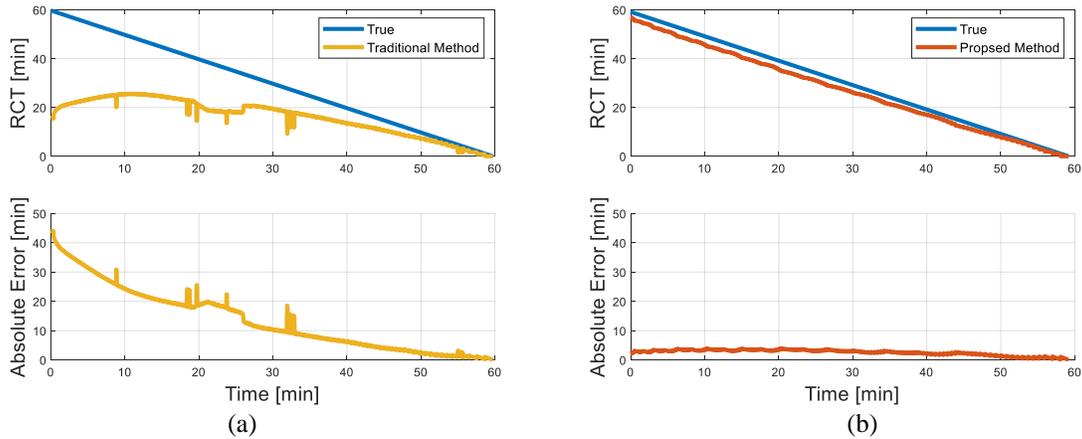
Fig. 16. The remaining charging time estimation results in the CV charging stage: (a) Traditional method, and (b) Proposed method with a well-trained resistance perdition model.

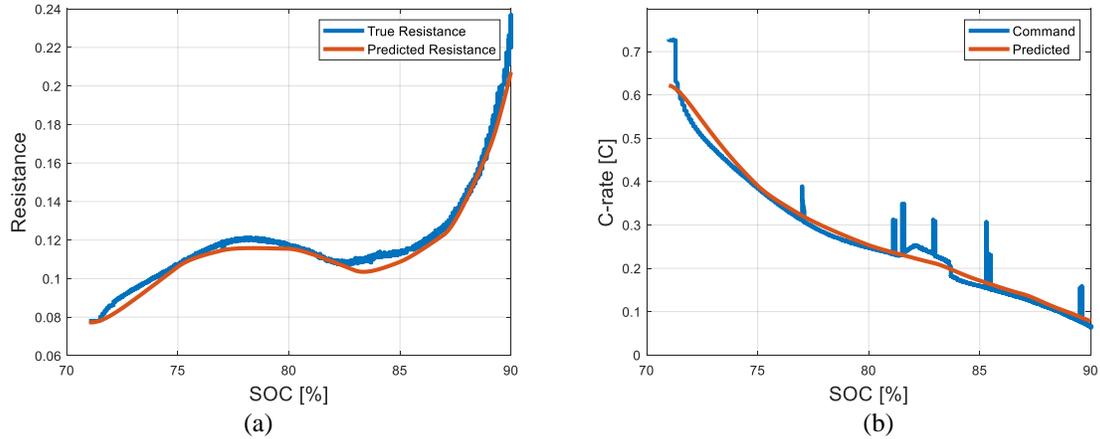
Fig. 17. The proposed method under the CV charging stage: (a) True and predicted resistances, and (b) Comparison of command and predicted C-rates.

An accurate battery resistance prediction model plays a crucial role in the RCT estimation in the CV stage. It also means that it is important to use online training or identification algorithms to capture the changing of battery resistances due to the battery degradation effect. To further demonstrate the performance of the online training used in the proposed method, the proposed method is tested by a new battery pack with the same design. In the test case, the initial parameters of the resistance prediction model are obtained from the previously tested battery pack, with a lower SOH value which means higher resistant values than the new one. As shown in Fig. 18 (a), with the wrong parameters set-up in the resistant prediction model, the maximum estimation error of RCT is 15 minutes at the first charging cycle. However, after new charging data is collected and the resistance prediction model is trained, as presented in Fig. 18 (b) and Fig. 18 (c), the RCT estimations of the new battery pack become more and more accurate. The maximum estimation error is 15 minutes in the initial charging cycle and is reduced to 8 minutes and 3 minutes in the second and third cycles, respectively.

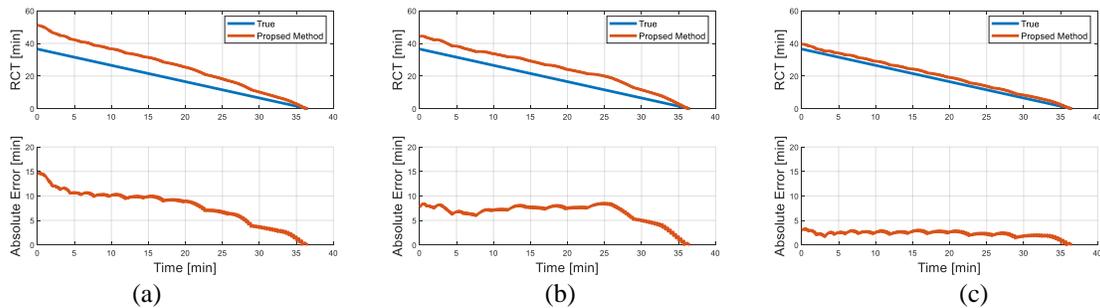
Fig. 18. The remaining charging time estimation results of the proposed method without prior knowledge about the battery pack: (a) first charging cycle, (b) second charging cycle, and (c) third charging cycle.

## 5. CONCLUSIONS

In this paper, a novel RCT estimation method is proposed and implemented in a production electric vehicle control system for performance validation. In the CC charging process, by taking advantage of updating charging accuracy in real-time, the RMSE of the proposed method is only 2.0165 minutes. In the CV charging process, the proposed method achieves accurate RCT estimation, which has an RMSE of only 2.735 minutes, with the assistance of a future current profile prediction. In addition, the impact factors on battery resistance in the CV stage are analyzed. Based on the experimental results, the battery temperature, starting SOC, and SOC are verified as the dominant factors impacting battery resistance. Therefore, they have been used as inputs to predict the resistance of a Li-ion battery in the CV charging process.

Besides applied to EVs, the proposed RCT estimation method can also be applied to other commercial and residential energy storage systems that use Li-ion batteries as their energy source, such as electric ships [35]. Meanwhile, the charging profile prediction capability of the proposed method can be extended to predict energy demand profiles of charging processes, which can assist energy management systems to improve the energy efficiency of microgrids [36].

In future studies, the application of machine learning is a promising area to continue making improvements in RCT estimation. Besides, by conducting the relationship between the capacity fading and resistance rising of Li-ion battery in the CV charging process, the resistance prediction model of the proposed method could be investigated for SOH and state-of-power estimation.